# Dynamic Shortening of Disorder Potentials in Anharmonic Halide Perovskites

Christian Gehrmann[1,2] and David A. Egger[1,2,*]

[1] *Institute of Theoretical Physics, University of Regensburg, 93040 Regensburg, Germany*
[2] *Department of Physics, Technical University of Munich, 85748 Garching, Germany*

**Abstract**

Halide perovskites are semiconductors that exhibit sharp optical absorption edges and small Urbach energies allowing for efficient collection of sunlight in thin-film photovoltaic devices. However, halide perovskites also exhibit large nuclear anharmonic effects and disorder, which is unusual for efficient optoelectronic materials and difficult to rationalize in view of the small Urbach energies that indicate a low amount of disorder. To address this important issue, the disorder potential induced for electronic states by the nuclear dynamics in various paradigmatic halide perovskites is studied with molecular dynamics and density functional theory. We find that the disorder potential is dynamically shortened due to the nuclear motions in the perovskite, such that it is short-range correlated, which is shown to lead to favorable distributions of band edge energies. This dynamic mechanism allows for sharp optical absorption edges and small Urbach energies, which are highly desired properties of any solar absorber material.

* Correspondence: david.egger@tum.de

Halide perovskites (HaPs) have emerged as semiconducting materials that are solution-processable and show an outstanding potential for device applications, notably for photovoltaics, where power-conversion efficiencies already approach those of silicon-based cells.[1–10] HaPs exhibit several physical properties that are key to their remarkable potential as a solar material, and in particular show a steep optical absorption rise that is required for an efficient capture of sunlight in a thin-film device.[11] Specifically, the Urbach energy, which quantifies the steepness of optical absorption rise, was found to be very small for HaPs at room temperature, on the order of 10-20 meV,[12] which is close to highly efficient inorganic solar materials such as bulk Si and GaAs. A small Urbach energy implies that the material is ordered, since disorder would induce tail states in the electronic structure broadening the optical absorption profile. Interestingly, however, ample experimental and theoretical evidence point to highly anharmonic nuclear motion and disorder being active in HaPs at room temperature, which also involves the ions in the crystal contributing to the frontier electronic band structure.[13–25] Such anharmonic effects are uncommon for efficient optoelectronic materials[26] and difficult to rationalize in view of the low Urbach energy. In particular, they can be suspected to result in a disordered potential for the electrons and holes in the crystal and, hence, in a high density of tail states and a broad optical absorption. Therefore, one must wonder how steep optical absorption edges, small Urbach energies and, hence, an efficient collection of sunlight can even be possible in thin-films of HaP crystals at room temperature.

Here, we use first-principles based molecular dynamics (MD) calculations to study and quantify the spatial correlations in the disorder potential for electrons and holes that is induced by thermal vibrations. We find that the *massive nuclear motions in HaPs lead to a dynamic shortening of the disorder potential for electrons and holes*, such that is confined to one atomic bond and becomes similar to the length-scales that were reported for classical inorganic semiconductors. Since the correlation length of the disorder potential is short-ranged, thermal nuclear motion disturbs the electronic states in only small volumes of the crystal. Therefore, this is expected to leads to a low density of disorder-induced tail states, which we demonstrate from first-principles calculations. This dynamic mechanism allows the here studied $CsPbBr_3$, $CsPbI_3$, and $MAPbI_3$ to exhibit narrow band-gap distributions, sharp optical absorption edges and small Urbach energies at elevated temperatures, which are highly desired properties of any solar absorber material used in thin-film devices.

*Results.*

**Spatial correlations in the disorder potential of CsPbBr3₃**

Our theoretical approach employs MD calculations based on density functional theory (DFT), which treat the nuclear anharmonicity to all orders in the Taylor expansion of the crystal potential and thereby allow for monitoring and quantifying its consequences for physical observables in a straightforward manner. Specifically, long MD trajectories at various temperatures were computed in order to characterize the consequences of the thermally-induced nuclear disorder potential for electronic states in the material, as illustrated in Fig. 1a. To this end, we firstly calculated the average electrostatic potential energy as

$$\bar{V}(x,y,z) = \frac{1}{N}\sum_{i=1}^{N} V_i(x,y,z), \quad (1)$$

where $V_i(x,y,z)$ is the electrostatic energy of the electrons for configuration $i$ along the MD trajectory, and $N$ is the total number of configurations we considered. Note that we chose $N = 30$ configurations, which were separated by 5 ps. We consider $\bar{V}$ to represent the average crystal potential for the electronic states. From this, we calculated the instantaneous disorder potential as the deviation from $\bar{V}$:

$$\Delta V_i(x,y,z) = V_i(x,y,z) - \bar{V}(x,y,z). \quad (2)$$

Finally, we computed the autocorrelation function of $\Delta V_i$ along the *y* coordinate as

$$C_i(x,\Delta y,z) = \frac{<\Delta V_i(x,y+\Delta y,z)\Delta V_i(x,y,z)>}{<\Delta V_i(x,y,z)\Delta V_i(x,y,z)>}, \quad (3)$$

and owing to the cubic symmetry of the crystal, calculated the average of it over all *x* and *z* values, which we denote as $C(\Delta y)$. Furthermore, the DFT-MD simulations allow for characterizing phonon quasiparticle properties by analyzing the velocity-autocorrelation function (VACF) in the basis of the harmonic spectrum (see Methods section for details). As a paradigmatic case, we choose to focus first on the all-inorganic HaP CsPbBr$_3$ in its cubic phase (Fig. 1a), and subsequently report results on CsPbI$_3$ and MAPbI$_3$ in order to test the influence of ionic composition.

Fig. 1b shows an exemplary charge-density response during the MD in CsPbBr$_3$, which was obtained as the difference of the density corresponding to a randomly-chosen nuclear configuration along the MD run at 425 K and the mean density along the MD trajectory. The calculated charge-density difference is such that it follows the nuclear displacements, and since all phonon modes are excited at the considered temperature (see below), it is essentially distributed throughout the entire simulation volume. What is perhaps more surprising is the fact that this charge-density difference appears highly disordered, i.e., it does not show any obvious symmetry properties or directional features.

Such a charge density response results in an instantaneous disorder potential for the electronic states in CsPbBr$_3$ via equation (2) that is spatially correlated by means of $C(\Delta y)$ given in equation (3), which we calculated self-consistently along the MD trajectory. $C(\Delta y)$ is shown in Fig. 1c for multiple instantaneous nuclear configurations along the MD trajectory, together with the average of these snapshots, for the three temperatures considered in the MD simulations. Remarkably, it is found that the correlation in the disorder potential is confined to very short ranges, since it vanishes on length scales of atomic bonds in CsPbBr$_3$. Furthermore, we find that at the three considered temperatures, the disorder potential shows a similar overall behavior. With this, we establish that the *spatial correlation in the disorder potential CsPbBr$_3$ is short-ranged*, such that it appears to be similar to the one present in classical inorganic semiconductors, which is correlated also only over atomic distances.[27]

**Vibrational characterization of CsPbBr$_3$**

The short-ranged spatial correlations for the electronic states in CsPbBr$_3$ are at first sight peculiar because the nuclear dynamics driving the disorder in HaPs were reported to be very different compared to the situation in classical inorganic semiconductors.[13–25] We characterize the key vibrational features of CsPbBr$_3$ in the following by first discussing results from the harmonic approximation. This description is valid for materials in which small nuclear displacements occur at finite temperature, as is the case in many of the classical inorganic semiconductors at room temperature. The harmonic phonon spectrum of CsPbBr$_3$ (see Fig. 2a) is found to contain low-energy features, especially due to Cs and Br motion, as was well as an intrinsic dynamic instability, which have been analyzed in previous theoretical work.[28,29] Next, we consider finite temperature vibrational properties of CsPbBr$_3$ obtained from a fully anharmonic treatment by means of DFT-based MD, analyzing the velocity autocorrelation function (VACF) as described in the Methods section. The structural instabilities, seen as imaginary features in the 0 K harmonic calculations of the phonon spectrum, vanish in the finite temperature dispersion relation (see Fig. 2a), which is expected. Fig. 2b shows the power spectrum of the VACF, which represents the vibrational density of states (VDOS), at three temperatures. Overall, while the vibrational features appear strongly broadened in the finite temperature spectra, one can still appreciate that they are rather similar to the harmonic result. Hence, theoretical data obtained from lattice dynamics calculations in the harmonic approximation are not generally in contrast to those obtained with a fully anharmonic treatment.

This motivates us to study the consequences of anharmonicity in greater detail, especially because the importance of it for optoelectronic properties was discussed in several previous articles.[13–25] Our approach is intrinsically based on first-principles information that is obtained at each timestep of the MD trajectory. Therefore, we are in a position to compute phonon quasiparticle properties and extract the vibrational lifetimes, which are key physical observables quantifying anharmonic nuclear effects. The phonon lifetimes, shown in Fig. 2c at different high symmetry points of the phonon branches as a function of frequency, are extremely short in $CsPbBr_3$, i.e., between 0.3 and 10 ps in a temperature range of 325 to 525 K. We also find that the phonon lifetime is shortest in the frequency range where the power spectrum exhibits the highest intensity, i.e., between 1-3 THz, which can be explained by considering that due to the presence of many phonons in this range the phonon-phonon interactions are strongest (cf. Fig. 2a, b and c). Table 1 establishes that both acoustic and optical phonons are very short-lived in $CsPbBr_3$, in agreement with recent experiments that studied the phonon lifetimes of acoustic modes in $MAPbI_3$.[25]

Fig. 2d shows histograms of the phonon lifetimes as a function of temperature. Some temperature-induced changes are indeed observed in the phonon lifetimes, e.g., certain modes do shift to lower lifetimes at higher temperatures, and the mean value of the distribution is lowered by ~0.3 ps from 325 to 525 K. Notably, however, the shape of the distribution remains largely similar and there is no apparent shift of the distribution to lower lifetimes with increase in temperature. Likewise, Table 1 shows the lower end of the range in the phonon lifetimes to remain largely constant (within a reasonable statistical error of ~80 fs) at different temperatures, while the upper end changes significantly. This finding can be explained by considering that in order to be physically meaningful, the period of a phonon with a given frequency presents a reasonable lower limit to its lifetime. We find that for CsPbBr$_3$, this limit has been reached already at 325 K, which again signals massive anharmonic effects in the nuclear displacements approaching the limit of a breakdown of the phonon picture. To put these findings in perspective, recall that in bulk Si the phonon lifetimes are two orders of magnitude higher for a similar range of temperature and frequency.[30] These findings strongly suggest that anharmonicity is an important phenomenon in the nuclear dynamics of HaPs, in agreement with previous studies on different HaPs.[13–25] It will now be shown that the mechanism underlying anharmonicity in HaPs renders the presence of a short-range correlated disorder potential particularly interesting.

**Resonant bonding and long-range effects in the disorder potential**

Resonant bonding was recently studied as a potential origin of the nuclear anharmonicity occurring in HaPs.[14] Such a mechanism was suggested 40 years ago for hybrid HaPs[31] and originates from the fact that multiple configurations of the *sp*-hybridized lead-halide framework are close in energy, as was discussed in previous work on HaPs[14,17] and other materials.[32,33] Importantly, resonant bonding causes long-range effects in the dynamical matrix,[32] which we illustrate with an example that was recently studied in the literature.[14] In Fig. 3a we plot the change in the charge density that is induced to the system upon displacing a single Pb atom showing that the response is clearly long-ranged in CsPbBr$_3$. Furthermore, in Fig. 3b we show that such a resonant bonding response would result in a *long-range correlated disorder potential* for electrons and holes in CsPbBr$_3$ by computing $C(\Delta y)$ for this specific case: the long-range effect in the density immediately translates into long-range correlations of the disorder potential that are induced upon the displacement, which exceed the ~5.8 Å unit cell of CsPbBr$_3$ by far. In contrast, the full MD calculations discussed above have established a *short-range correlated disorder potential* that was confined to distances of one atomic bond.

Resolving the discrepancy between long-range disorder effects implied by a resonant bonding mechanism and the appearance of short-ranged correlated disorder in the actual MD calculations is important. While the role of disorder for properties of solar materials has been long known,[34] revealing the origin and consequences of the short-range correlated disorder present in HaPs will provide a fresh view on the interrelation of nuclear dynamics and optoelectronic properties in these materials. Consider the case of long-range correlated disorder (see Fig. 4a), defined as being correlated across several atomic bond distances, which would perturb the electronic states in large fractions of the crystal volume. A macroscopic sample of such a material would therefore contain a high density of tail states if long-range correlated disorder was indeed active. The Urbach energy would then be large, since it scales with the square of the correlation length of the disorder potential.[27] In contrast, the type of short-range correlated disorder we found in $CsPbBr_3$, which is correlated only across distances on the order of one atomic bond, perturbs the electronic states in only small fractions of the crystal volume (see Fig. 4b). Hence, in a macroscopic sample of $CsPbBr_3$, different regions of the crystal are perturbed electronically only independently of one another by the nuclear dynamics. Importantly, while locally and temporally the nuclear configuration may generally be strongly perturbed from the ideal lattice, this effect is expected to induce tail states in only a relatively small crystal volume. Therefore, a large crystal sample of $CsPbBr_3$ is expected to *on average containing only a small density of disorder-induced tail states*, the distribution of available band gaps will be narrow, and the optical absorption edge will be sharp. This is a prerequisite for the small Urbach energy that was recorded experimentally for this[35] and other HaPs,[12] which is a highly desired property of any solar material. It also implies that the impact of thermal nuclear disorder on the optoelectronic properties of Cs-based HaPs is such that the optical absorption onset is still sharp. Hence, an efficient collection of sunlight in thin-film devices should be possible given the intrinsic properties of these materials. The mechanism underlying the short-range disorder potential, however, is expected to be different in HaPs compared to classical inorganic semiconductors such as bulk Si and GaAs. For Si, it was shown that even for a statically displaced structure the response is short-ranged,[14] i.e., in Si it does not require any nuclear dynamics to find a short-range correlated disorder potential.

**Dynamic shortening of spatial correlations and narrowing of dynamical band-edge distributions**

We now explore the impact of nuclear dynamics on the spatial correlation in the disorder potential of electrons and holes in more detail. To this end, we consider several relevant constrained scenarios of nuclear motion that occur in the full MD of $CsPbBr_3$ at T=425 K. For each considered case, we calculate how it impacts the spatial correlations in the disorder potential of the material. Fig. 5a shows data for $C(\Delta y)$ corresponding to three cases: (i) configurations of Pb atoms chosen from the MD trajectory, but *Cs and Br atoms being fixed to their ideal lattice positions*; (ii) configurations of Pb and Br atoms chosen from the MD trajectory, but *only Cs atoms being fixed to their ideal lattice positions*; and (iii) configurations of Pb and Cs atoms chosen from the MD trajectory, but *only Br atoms being fixed to their ideal lattice positions*. It is found that Pb displacements induce a disorder potential for the electronic states showing a spatial correlation that significantly exceeds atomic distances, in line with the expectation borne from the resonant bonding mechanism discussed above. Most importantly, the correlation length of the *disorder potential for electrons and holes is massively shortened by activating either Cs or Br displacements*. Cs displacements already cause a substantial reduction of the correlation length, but it is the presence of displaced Br atoms that reduces spatial correlations in the disorder potential such that it becomes similar to the result obtained in fully-unconstrained MD (see Fig. 1c). Hence, the disorder potential for electrons and holes is dynamically shortened by the large nuclear motions of Cs and especially Br at elevated temperatures in $CsPbBr_3$.

A manifestation of electron-phonon interactions in $CsPbBr_3$, namely the impact of spatial correlations in the disorder potential on the density of tail states (see Fig. 4), can now be tested explicitly. The investigation of this effect is based on the above finding that when only Pb displacements are active, the correlation length was found to be longer than in the full MD calculation (cf. Figs. 5a and 1c). In Fig. 5b, we therefore report normalized histograms of DFT-calculated band-edge energies that occur during constrained MD calculations in which only Pb displacements were considered, and those of the full MD run (see Methods section for details). First, all finite temperature distributions for the valence band maximum (VBM) and conduction band minimum (CBM) are broadened compared to the static DFT result. This is expected because multiple nuclear configurations are sampled at elevated temperature, which affects the orbital overlap in the perovskite lattice and leads to fluctuations in the instantaneous VBM and CBM energies. Second, considering the case where only Pb displacements are active (blue curve in Fig. 5b), it can be seen that the distribution of CBM energies is wider than the one of VBM energies. This finding can be explained by the anti-bonding character of the valence band[36] and the fact that the long-range nature of resonant bonding is mostly due to Pb *p*-states, which contribute strongly to the CBM states.[6] Hence, the effect of just Pb displacements being active in CsPbBr$_3$ leads to tails in the CBM energy distribution, which is again expected. Third and most important, when we now consider the result of the full MD calculation (red-filled curve in Fig. 5b), we find that the distribution of CBM energies becomes much narrower compared to the case where only Pb displacements were active. This finding reveals that the shortening of the disorder potential (cf. Figs. 5a and 1c) manifests itself in a reduction of the density of tails in the CBM energy distribution.

Therefore, we have explicitly established the connection between the short-range correlated disorder potential and favorable distribution of electronic states in CsPbBr$_3$. The finding implies that in absence of defect states and other static perturbations of the crystal, such a short-range correlated disorder potential will result in the recovery of one of the hallmark optoelectronic properties of HaPs, namely a sharp optical absorption edge and small Urbach energy at elevated temperatures. These are key properties of any solar material since they allow for efficient collection of sunlight in thin-film crystalline materials.

**Influence of ionic composition**

A variety of ionic compositions are compatible with the ABX$_3$ stoichiometry of HaP crystals. Several recent experimental and theoretical data showed strong anharmonic and disorder effects for HaPs compounds containing various A-site cations and X-site anions,[13–25] which nevertheless exhibit favorable optoelectronic properties including a small Urbach energy. Therefore, it is interesting to examine whether the above-described short-ranged correlated disorder for the electronic states in CsPbBr$_3$ is present also in alternative HaP compounds. To this end, in a first step we exchange bromine by iodine, and then cesium by the organic methylammonium (MA) molecule, to investigate the disorder potential in the CsPbI$_3$ and MAPbI$_3$ crystal, respectively.

In Fig. 6a, we show $C(\Delta y)$ of CsPbI$_3$, calculated for multiple instantaneous nuclear configurations along the MD trajectory (T=425 K) together with their average. It is found that the disorder potential is markedly short-ranged, since it vanishes on the length scale of atomic bonds in CsPbI$_3$, similar to the situation in CsPbBr$_3$ (see Fig. 1c). To investigate the effect of the nuclear dynamics on the disorder in CsPbI$_3$, we perform the same procedure that was described above for CsPbBr$_3$ and consider constrained scenarios of nuclear motion. The results (see Fig. 6b) show that the mechanism of a dynamic shortening of the disorder potential is indeed active also in CsPbI$_3$. In particular, the longer-range disorder induced by Pb displacements is shortened strongly by activating Cs and especially I displacements. While the mechanism of dynamically shortening the disorder in CsPbI$_3$ is thus remarkably similar to the situation in CsPbBr$_3$ (cf. Fig. 5a and 6b), there is also one interesting difference. The partially negative correlation induced by activating only Pb displacements in CsPbBr$_3$, at a distance of approximately two lattice constants (see Fig. 5a), is hardly visible for the averaged curve in case of CsPbI$_3$ (see Fig. 6b). This suggests that a static lattice of iodine atoms provides more screening of the disorder potential than the one of bromines, which is reasonable given the larger atomic polarizability of the former. Similar to the case of CsPbBr$_3$, in Fig. 6c we demonstrate explicitly that the shortening of the disorder potential corresponds to reducing the spread of CBM energies in CsPbI$_3$. Note that an interesting difference compared to CsPbBr$_3$ is that Pb displacement induce a more symmetric tailing of CBM energies.

Finally, we consider the hybrid organic-inorganic HaP MAPbI$_3$, which is interesting because first of all it was recently demonstrated that long-range effects due to a resonant bonding mechanism are active in this material.[14] Furthermore, the anisotropy and dipole moment of the MA molecule could provide means for both, additional long-range order due to oriented dipoles, as well as additional disorder and screening due to disarranged dipoles. In Fig. 7 we show $C(\Delta y)$ of MAPbI$_3$ calculated by combining trajectories obtained in force-field MD with DFT calculations, as described in the Methods section. Our results for MAPbI$_3$ establish that also in this compound the disorder potential due to thermally activated nuclear displacements is short-ranged. This is in line with the established fact that at higher temperatures, the orientation of MA is largely disordered. Note that it is the lacking dipolar order in MAPbI$_3$, which renders the subsequent analysis of constrained MD scenarios shown above for the Cs-based compounds very challenging. With this, we conclude that the dynamically short-ranged disorder potential is present in all considered HaP compounds, which is relevant for their favorable optoelectronic characteristics.

*Discussion.*

By means of MD calculations we have reported a theoretical analysis of the disorder potential that is induced by thermal vibrations in HaPs in order to characterize spatial correlation properties. The latter are key for understanding the microscopic origin of the small Urbach energies that were recorded experimentally for HaP compounds, which enable these materials to be used as efficient solar absorbers in thin-film devices. We found that the correlation length of the disorder potential for electrons and holes is markedly short-ranged in the considered compounds $CsPbBr_3$, $CsPbI_3$ and $MAPbI_3$, since it was found to be confined to distances of only one atomic bond, which is similar to the length-scales that were reported for bulk Si and GaAs.[27] Notably, this is in agreement with previous findings, which studied the nuclear motion contributing to the central peak in the $CsPbBr_3$ Raman spectrum[16] and quantified the correlation length of the disorder potential in $MAPbI_3$ using classical MD.[37] The presence of such a short-range correlated disorder potential in HaPs is peculiar since strongly anharmonic nuclear motions and disorder effects were reported for these systems at room temperature,[13–25] which may be suspected to perturb the electronic disorder in a profound way.

In search for the origin of the short-range correlated disorder potential in HaPs, we considered the microscopic modulations of the electrostatic potential in the material that are induced by the nuclear displacements. Our data showed that a resonant bonding mechanism, which was suggested to be active in HaPs already 40 years ago,[31] is present in these compounds. This result is in strong contrast to the case of classical inorganic semiconductors such as bulk Si or GaAs, but similar to previous findings related to HaPs[14,17] and thermoelectric compounds.[32,33] Resonant bonding in HaPs stems from the fact that multiple nuclear configurations of the *sp*-hybridized lead-halide framework are energetically very close, such that nuclear displacements cause long-range effects in the charge density of the system. Since we have also demonstrated that a resonant bonding mechanism would establish long-range correlated disorder potentials for electrons and holes, one must ask how it is possible that HaPs dynamically exhibit a disorder potential that is short-ranged.

To address this question, we could show that it is the massive nuclear motion of the A-site cation and especially the halides leading to a dynamic shortening of the disorder potential induced for electrons and holes. The results from our first-principles calculations explicitly established that the dynamically short-ranged disorder potential leads to favorable dynamic distributions of band edge energies. We conclude that it is this dynamic effect which leads to large crystalline samples of HaPs containing only few disorder-induced tail states. This implies sharp optical absorption edges and low Urbach energies, which are key physical properties of any solar absorber material since they allow for efficient collection of sunlight in thin-film devices.

*Methods.*

**Density functional theory (DFT) calculations**

DFT calculations were performed with the plane-wave code VASP,[38] using the projector augmented wave (PAW) method to treat core-valence interactions.[39] Unless stated otherwise, we employed the "normal" version of the code-supplied PAW potentials. Exchange-correlation was described with the PBE functional,[40] augmented by dispersive corrections computed in the Tkatchenko-Scheffler scheme,[41] which was shown to provide an accurate description of static and dynamic structural properties of HaPs.[42] Unless stated otherwise, an energy threshold of $10^{-8}$ eV, a Γ-centered *k*-point grid of 6x6x6, and a plane-wave cutoff energy set to 500 eV were used. The cubic lattice structure of all compounds were optimized with these settings such that residual forces were below $10^{-3}$ eV/Å, for which improved convergence thresholds did not result in any significant changes. This structure was used in our subsequent calculations. Structural representations of $CsPbBr_3$ were visualized using the VESTA program.[43]

**Lattice dynamics**

To obtain the phonon dispersion relation and vibrational density of states, lattice dynamics calculations were performed using the finite displacement method as implemented in the phonopy package.[44] In these calculations, a 2x2x2 supercell of the optimized $CsPbBr_3$ structure was used, and all numerical parameters were kept as before, applying a *k*-point grid that was reduced according to the enlarged cell size.

**DFT-based molecular dynamics**

First-principles molecular dynamics (MD) calculations were performed using a canonical (*NVT*) ensemble with a Nosé-Hoover thermostat, as implemented in the VASP code, employing a timestep of 8 fs. We have chosen a larger 4x4x2 (160 atom) supercell in the MD simulations in order to improve the statistical sampling of the nuclear dynamics. For making these simulations computationally tractable, more efficient numerical settings were employed: the "GW" PAW potentials were used, since this improved the numerical convergence of the self-consistent calculations of the MD runs, together with a plane-wave cutoff energy of 250 eV, an energy threshold of $10^{-6}$ eV, and a single $k$ point. We verified that the latter was sufficiently accurate, by monitoring the power spectrum of the velocity autocorrelation function calculated with more $k$ points. The system was equilibrated for at least 5 ps at each temperature, and the subsequent nuclear dynamics were analyzed along trajectories of 150 ps, which is more than ten times longer than the longest calculated phonon lifetime.

**Force-field molecular dynamics**

Force-field MD calculations were performed on a 4×4×2 supercell of cubic $MAPbI_3$ (384 atoms) with the LAMMPS code,[45] applying the force field by Mattoni et al.[46] We used a canonical (*NVT*) ensemble with a Nosé-Hoover thermostat, employing a timestep of 0.5 fs. The system was equilibrated for 1 ns followed by a production run of 150 ps.

**Phonon-quasiparticle properties**

Phonon-quasiparticle properties of $CsPbBr_3$ were calculated using the dynaphopy package.[47] Specifically, the MD-calculated velocity autocorrelation functions were analyzed by means of a projection of them onto the harmonic modes which were calculated as described above. Fitting the power spectrum of these mode-resolved projections with Lorentzian functions provided well-defined frequencies and lifetimes of each phonon mode as a function of the phonon wavevector, **q**, at high symmetry points. Note that the acoustic phonons at the Γ-point were not included in this analysis, since these are zero-frequency modes. The finite-temperature phonon dispersion was then obtained by calculating renormalized force constants that correspond to the quasiparticle phonon frequencies at the high symmetry points, which provided an updated dynamical matrix that was then interpolated in **q**, akin to lattice dynamic calculations. Note that the time-frequency transform was achieved by means of standard Fast-Fourier-transform. Further theoretical details can be found in refs. 47 and 48.

**Disorder potential and band-edge histogram calculations**

To calculate the disorder potential and band-edge histograms, we selected instantaneous nuclear configurations along the MD trajectories and computed the electrostatic potential energy and electronic structure of the system self-consistently with DFT. To enhance the accuracy of these calculations, compared to the DFT-MD simulations we increased the cutoff energy to 500 eV and used a *k*-point grid of 1x1x2 in accordance with the real-space dimensions of the supercell. The autocorrelation function, which we denoted as $C(\Delta y)$, was calculated as described above (see equations (1)-(3)). $C(\Delta y)$ shown in Fig. 3b was obtained in an equivalent way, namely by calculating the disorder potential as the deviation of the electrostatic potential energy of the considered nuclear configuration (one Pb atom displaced by 5% of the primitive lattice constant) from the potential energy of the ideal one. In the latter calculations, we used a 5x5x2 supercell to minimize boundary effects that are induced by the long-range nature of the density response. $C(\Delta y)$ shown in Fig. 5a and 6b was also obtained equivalently, namely by calculating the disorder potential as the deviation of the electrostatic potential energy of the considered nuclear configuration (see Fig. 5a and 6b) from the potential energy of the one averaged for all considered configurations. To compute the band-edge distributions, 90 structures separated by 200 steps or 1.6 ps in the MD were considered. The normalization of the histograms was performed for the valence band maximum and conduction band minimum of each type of MD calculation separately. To sacrifice detail for clarity, we have shown the histograms as a function of a dimensionless parameter.

*Acknowledgements.* We thank Omer Yaffe and David Cahen (both Weizmann Institute of Science) for fruitful discussions. Funding provided by the Alexander von Humboldt-Foundation in the framework of the Sofja Kovalevskaja Award, endowed by the German Federal Ministry of Education and Research, is acknowledged. The authors gratefully acknowledge the Gauss Centre for Supercomputing e.V. for funding this project by providing computing time through the John von Neumann Institute for Computing on the GCS Supercomputer JUWELS at Jülich Supercomputing Centre.

*Author contributions.* C.G. performed the theoretical calculations and analyzed the data. D.A.E. conceived and supervised the project. C.G. and D.A.E. interpreted the results and wrote the manuscript.

**Table 1**

Table 1: Range for the lifetimes of acoustic and optical phonons in $CsPbBr_3$, calculated for different temperatures.

| Temperature (K) | Acoustic modes (ps) | Optical modes (ps) |
|---|---|---|
| 325 | 0.5 – 9.5 | 0.3 – 4.6 |
| 425 | 0.4 – 6.7 | 0.3 – 4.4 |
| 525 | 0.5 – 5.4 | 0.3 – 2.8 |

**Figure 1**

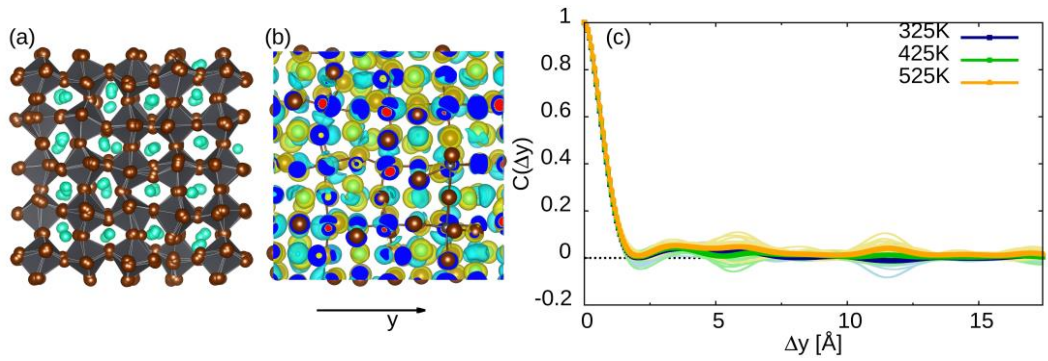

Fig. 1: Characterization of the disorder potential for electronic states in $CsPbBr_3$. a) Visualization of fully anharmonic dynamics in $CsPbBr_3$, which include phonon-phonon interactions in the finite-temperature description of the dynamic distortions present in the crystal. Cs atoms are shown in cyan, Pb atoms in gray, and Br atoms in brown color; the latter two species form gray-colored octahedra. b) Iso-surface representation of the charge-density difference that is induced in the crystal at an instantaneous configuration along the MD trajectory (T=425 K) with respect to the mean density. It can be seen that there is no apparent symmetric or directional response in the charge density. c) Autocorrelation function of the change in the electrostatic potential energy, which represents a disorder potential for the electronic states and was calculated for the direction indicated in panel b, along the MD simulation at different temperatures. The thin curves show sample snapshots taken along the MD trajectory, and the thick curves their averages. The disorder potential in $CsPbBr_3$ is confined dynamically to very short ranges of nearest-neighbor atomic distances.

**Figure 2**

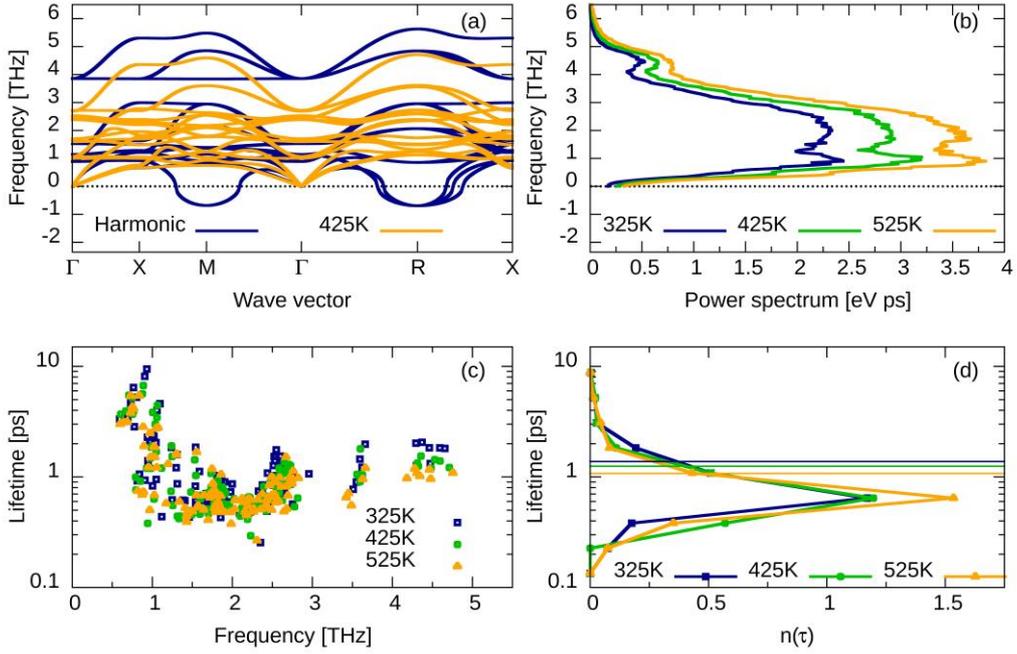

Fig. 2: Vibrational properties and anharmonicity in CsPbBr$_3$. a) harmonic (blue) and finite-temperature renormalized phonon dispersion at T = 425K (orange), as obtained from lattice dynamics calculations in the harmonic approximation and fully anharmonic MD, respectively. b) Power spectrum of the velocity autocorrelation function, as obtained from fully anharmonic MD, calculated at three temperatures. d) Phonon lifetime as a function of the renormalized phonon frequency, shown at three temperatures. d) Histogram of the phonon lifetimes, τ, shown for each of the three considered temperatures; see Table 1 for further details. The vertical lines indicate the mean value of the distribution at each temperature, which are 1.38 ps, 1.25 ps, and 1.08 ps at 325 K, 425 K, and 525 K, respectively.

**Figure 3**

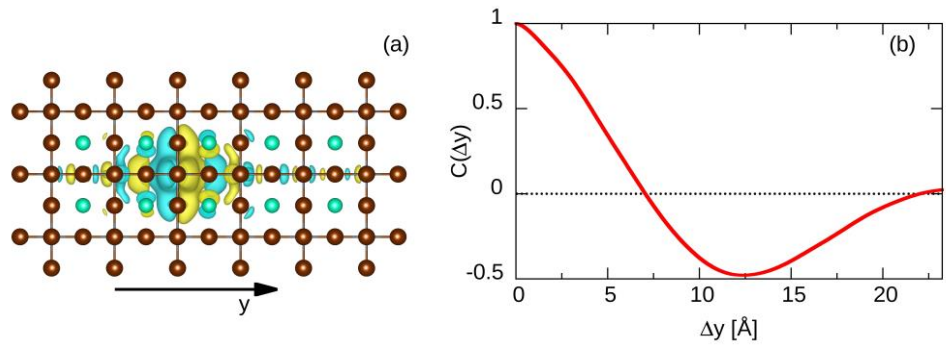

Fig. 3: Resonant bonding in CsPbBr$_3$. a) Iso-surface representation of the charge-density difference that is induced in the crystal upon displacing a Pb atom, which shows a long-range behavior. b) Autocorrelation function of the change in the electrostatic potential energy calculated for the direction indicated in panel a. It represents a disorder potential for the electronic states and also shows a long-range behavior.

**Figure 4**

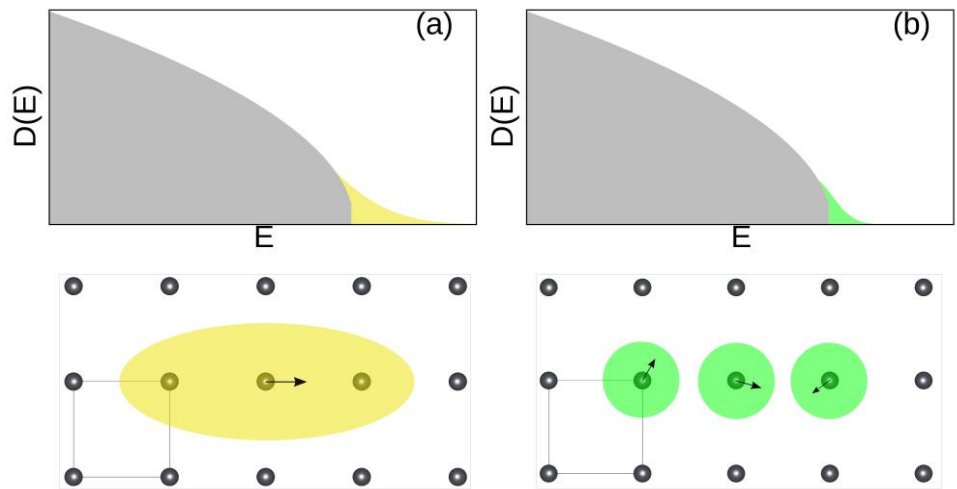

Fig. 4: Sketch of the impact of different kinds of spatially correlated disorder potentials on the electronic states in a crystal. a) Long-range correlated disorder induced by a single displacement (signified by the arrow) that leads to a long-range disorder potential, illustrated as a yellow disorder domain. It implies a larger amount of tail states in the electronic density of states, D(E), since larger parts of the crystal volume are perturbed by each disordered nuclear configuration. b) Short-range correlated disorder is illustrated by green disorder domains in the lower panels that are characterized by uncorrelated displacements (signified by the arrows). It leads to a small amount of tail states in D(E), since only small parts of the crystal volume are perturbed by each disordered nuclear configuration.

**Figure 5**

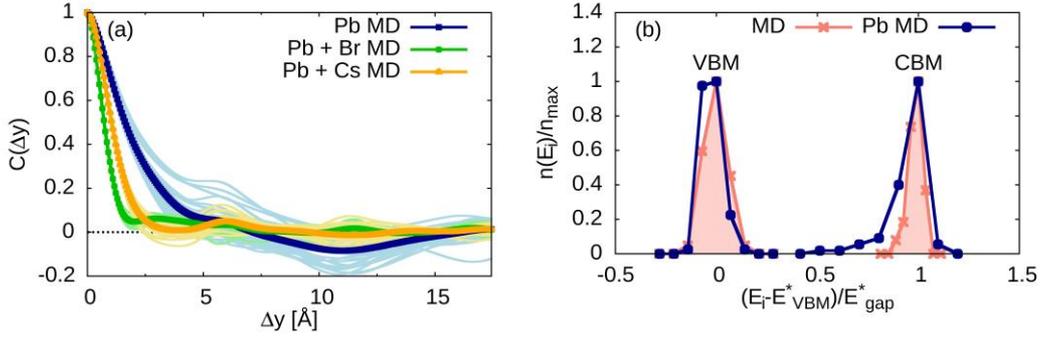

Fig. 5: Dynamic shortening of the disorder potential and narrowing of dynamical band-edge distributions in CsPbBr$_3$. a) Autocorrelation function of the disorder potential corresponding to three scenarios, where the configurations of Pb atoms were always chosen from the MD simulation (T=425 K), and: (i) Cs and Br atoms were fixed to their ideal lattice positions (blue curve); (ii) configurations of Br atoms were also chosen from the MD trajectory, but Cs atoms were fixed to their ideal lattice positions (green curve); and (iii) configurations of Cs atoms were also chosen from the MD trajectory, but Br atoms were fixed to their ideal lattice positions (orange curve). The thin curves show sample snapshots taken along the MD trajectory (T=425 K), and the thick curves their averages. The longer correlation length in the disorder potential induced by the Pb displacements is dynamically shortened by the Cs and especially the Br displacements. b) Normalized histograms of the instantaneous energies, E$_i$, of the valence band maximum (VBM) and conduction band minimum (CBM), calculated along two relevant MD scenarios: constrained configurations for which Pb atoms were chosen from the MD simulation (T=425 K), but all other atoms were fixed to their ideal lattice positions (blue curve with dots); and configurations for which all atoms are displaced according to the MD calculation at T=425 K (red curve with stars). $E^*_{VBM/CBM}$ denotes the VBM/CBM energy with the highest occurrence, $n_{max}$, which is given by $n_{max} = n(E^*_{VBM/CBM})$, and $E^*_{gap} = E^*_{CBM} - E^*_{VBM}$; see Methods section for details. The spread in the CBM energy distribution due to the displaced Pb atoms is reduced by the presence of other displacements occurring in the material.

**Figure 6**

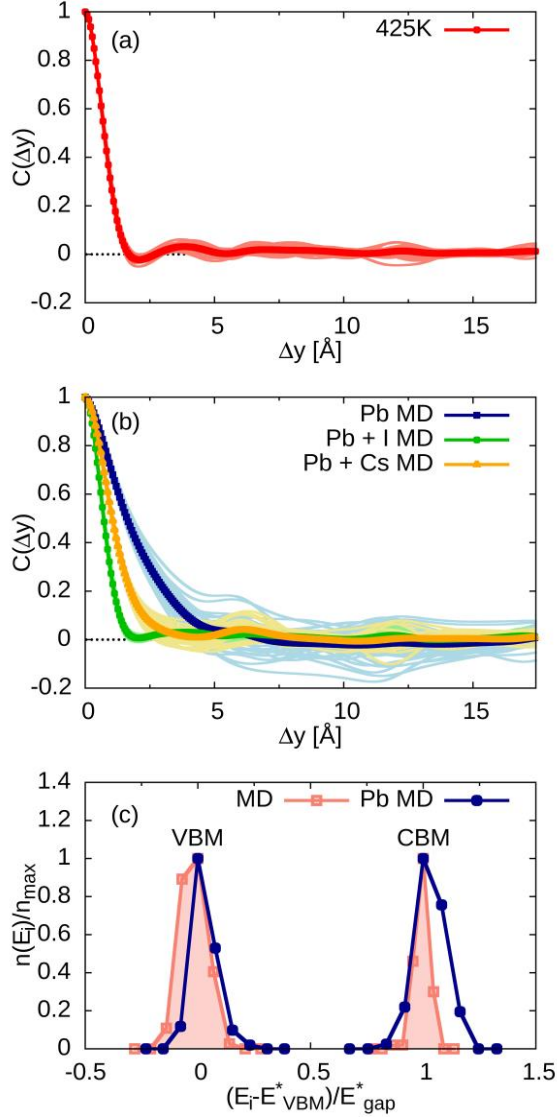

Fig. 6: Characterization of the disorder potential for electronic states in $CsPbI_3$. a) Autocorrelation function of the disorder potential, where the thin curves show sample snapshots taken along the MD trajectory, and the thick curves shows their average. The disorder potential in $CsPbI_3$ is also confined to very short ranges. b) Autocorrelation function of the disorder potential in $CsPbI_3$ corresponding to the three equivalent scenarios shown in Fig. 5a, where the thin curves show sample snapshots taken along the MD trajectory (T=425 K), and the thick curves their averages. Also in $CsPbI_3$ the correlation length in the disorder potential is dynamically shortened by the Cs and especially the I displacements. c) Normalized histograms of the instantaneous VBM and CBM energies calculated equivalently to the data shown Fig. 5b. In $CsPbI_3$ the spread in the CBM energy distribution due to displaced Pb atoms is also reduced by the presence of the other displacements in the material.

**Figure 7**

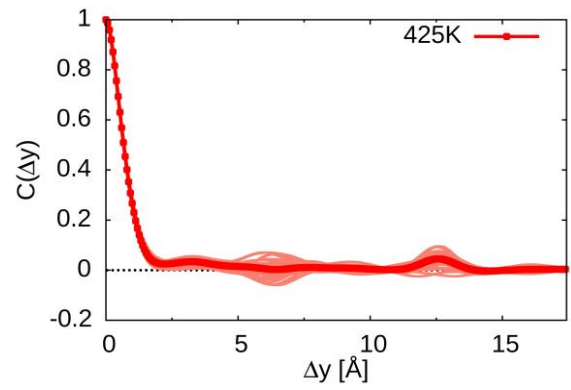

Fig. 7: Characterization of the disorder potential for electronic states in MAPbI$_3$. Autocorrelation function of the disorder potential, where the thin curves show sample snapshots taken along the force-field MD trajectory, and the thick curves shows their average. The disorder potential in MAPbI$_3$ is also confined to very short ranges.